\documentclass[12pt,twoside]{article}
\usepackage[T1]{fontenc}
\usepackage[latin1]{inputenc}
\usepackage{times}
\usepackage{graphicx}
\usepackage{a4wide}
\usepackage{listings}
 \usepackage{amsmath}
\usepackage{amssymb}

\begin{document}

\section*{Design considerations for the Micro Vertex Detector of the Compressed Baryonic Matter 
  experiment\footnote{Presented at the 17th International Workshop on Vertex detectors
    July 28 - 1 August  2008
    Uto Island, Sweden
}}

\thispagestyle{empty}

\begin{raggedright}

\markboth{M. Deveaux {\it et al.}}
{Design considerations for the MVD of the CBM experiment}

M.~Deveaux$^1$, S.~Amar-Youcef$^1$, C.~Dritsa$^{1,2,3}$, I.~Fr\"ohlich$^1$,
C.~M\"untz$^{1,2}$, S.~Seddiki$^{1,3}$, J.~Stroth$^{1,2}$, T.~Tischler$^1$ and 
C.~Trageser$^1$
\vspace{1cm}\\
\hspace{-0.4cm}\makebox[0.3cm][r]{$^{1}$}
Institut f\"{u}r Kernphysik, Johann Wolfgang Goethe-Universit\"{a}t, 60438 ~Frankfurt, Germany\\
\hspace{-0.4cm}\makebox[0.3cm][r]{$^{2}$}
Gesellschaft f\"{u}r Schwerionenforschung mbH, 64291~Darmstadt, Germany\\
\hspace{-0.4cm}\makebox[0.3cm][r]{$^{3}$}
Institut Pluridisciplinaire Hubert Curien (IPHC),
67037 Strasbourg, France
\end{raggedright}
\vspace{1cm}
\begin{center}
\textbf{Abstract}
\vspace{0.5cm}\\
\begin{minipage}[c]{0.9\columnwidth}
The CBM experiment will investigate heavy-ion collisions at beam
energies from 8 to 45 AGeV at the future accelerator facility
FAIR. The goal of the experiment is to study the QCD phase diagram in
the region of moderate temperatures and highest net-baryon densities
in search of the first-order phase transition from confined to
deconfined matter at the QCD critical point.  To do so, CBM aims to
measure rare hadronic, leptonic and photonic probes among them open
charm.  In order to reject the rich background generated by the heavy
ion collisions, a micro vertex detector (MVD) providing an
unprecedented combination of high rate capability and radiation
hardness, very light material budget and excellent granularity is
required.  In this work, we will present and discuss the concept of
this detector.
\end{minipage}

\end{center}
 
\section{Introduction}
\subsection{CBM, a FAIR experiment}
	
The FAIR facility at GSI \cite{FAIR} will offer unique possibilities
for the investigation of the QCD phase diagram in the regime of large
net-baryon densities, besides serving a variety of other fields of
physics with i) anti-proton beams for hadron physics, ii) radioactive
beams for nuclear structure physics, and iii) highly pulsed ion beams
for plasma physics. For the nuclear collision program, a synchrotron
with 300 Tm bending power (SIS-300) will deliver fully stripped heavy
ion beams up to uranium with intensities of up to $2 \cdot 10^9$ per
second at beam energies from 8 to 35 AGeV. Lighter ions (Z/A = 0.5)
can be accelerated up to 45 AGeV, while proton beams will be available
up to 90 GeV. The unprecedented beam intensities will allow studying
extremely rare probes with high precision but also constitute a high
challenge for detectors and electronics.  The CBM (Compressed Baryonic
Matter) experiment \cite{CBM} will be a next-generation fixed target
detector to be operated at the FAIR heavy-ion synchrotron SIS-300. It
is designed to measure hadronic, leptonic and photonic probes in a
large acceptance and at the extreme interaction rates offered by the
accelerator. CBM aims at a systematic investigation of $A+A$, $p+A$ and
$p+p$ collisions, in terms of collision energy ($\sqrt{\rm S_{NN}} =
4.5-9.3~ \rm GeV$ for heavy nuclei) and system size, with high
precision and statistics. In contrast to the low-energy programs at
the RHIC and the SPS, which due to low collision rates will focus on
bulk particle production, CBM will put special emphasis on the
measurement of extremely rare probes which have not been accessible by
previous heavy-ion experiments at the AGS and the SPS.

The observables to be covered by CBM include multiplicities, phase
space distributions and the flow of strange, multi-strange (K,
$\phi$,$\Lambda$, $\Xi$, $\Omega$) and charmed hadrons (D, D$_{\rm
  S}$,$\Lambda_{\rm C}$). Short lived vector mesons and charmonium
states will be investigated via their di-leptonic decay. The
measurements on charmonium states together with open charm
measurements will allow a comprehensive study of charm production near
the production threshold. Signatures of the critical point will be
looked for in event-by-event fluctuations of the quantities like
particle yield ratios, charged multiplicity or average $\rm p_t$.

The envisaged measurements of rare probes calls for an unique
instrument providing simultaneously an outstanding rate capability and
precision. Combining both is the central design challenge of the CBM
experiment. Our global design concept is discussed in
\cite{VolkerPaper}. This work will concentrate on the measurement of
open charm particles and on the Micro Vertex Detector of CBM.  To do
so, in section \ref{SectionGlobalGeometry}, we will introduce the
global geometry of the the MVD of CBM. Hereafter, in section
\ref{RequirementsMVD}, we will first discuss in detail the
requirements on the detector system. In section
\ref{ChapterTechnologicalConstraints}, we will motivate our technology
choices and discuss the constraints arising from the features and
limits of our guide line technology, which are CMOS Monolithic Active
Pixel Sensors. Knowing those constraints, we will introduce the design
of our detector ladders and estimate its material budget (section
\ref{SectionLadderConcept}). Finally, in section
\ref{SectionSimulation}, we will propose a running scenario and show
some preliminary simulation results of the physics performances of CBM
in the field of open charm reconstruction.

\subsection{The CBM Micro Vertex Detector (MVD)}
\label{SectionGlobalGeometry}
\begin{figure}[tb]\hspace{0.5cm}
  \begin{minipage}[c]{7.5cm}
    \includegraphics[width=7cm]{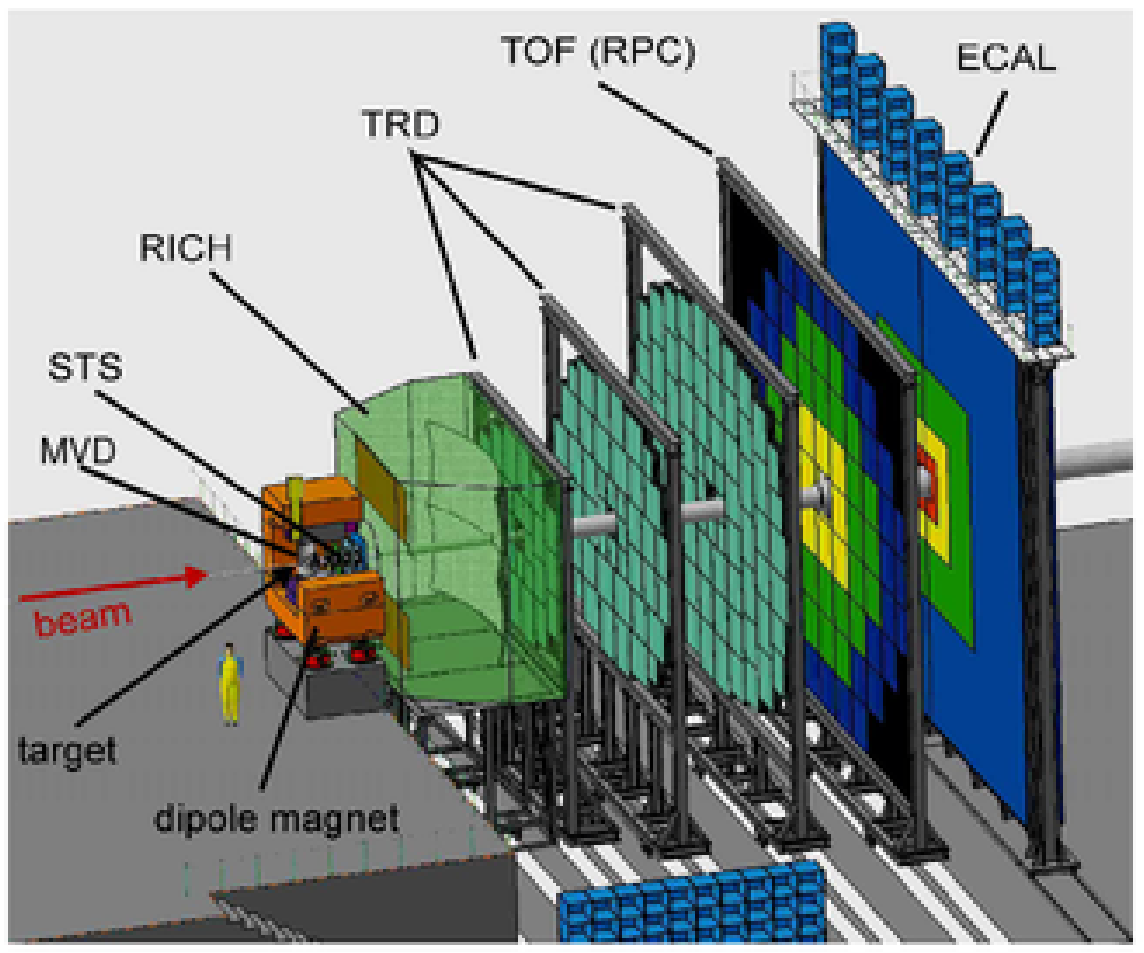}
  \end{minipage}
  \begin{minipage}[c]{7.5cm}
    \includegraphics[width=7cm]{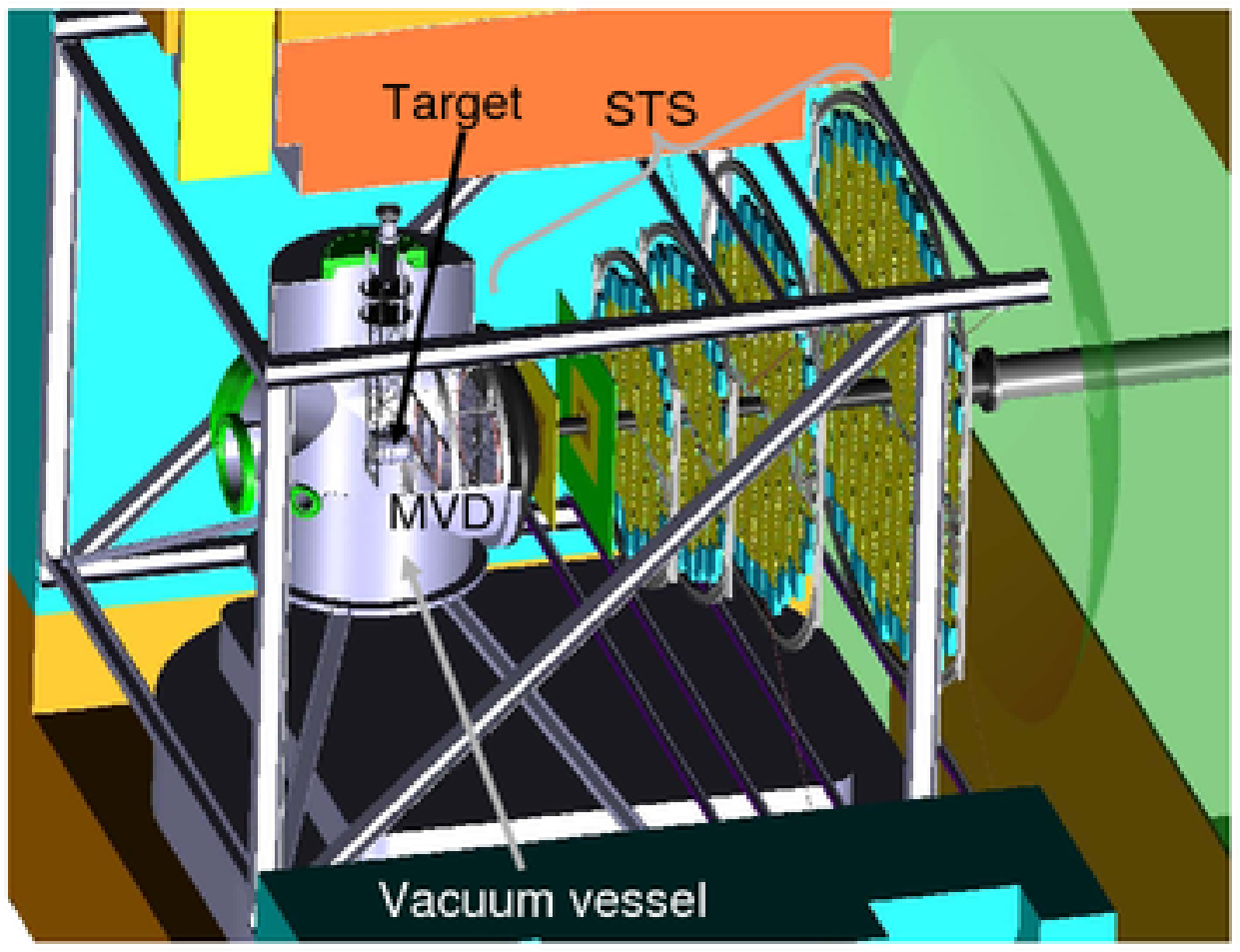}
  \end{minipage} 
  \caption[Significance Estimate] {\it Artistic view of the global
    layout of CBM (left) and a zoom into the MVD and STS (right).\\ }
  \label{GlobalCBM}
  
\end{figure}

The CBM experiment is currently planned with two configurations among
which one is optimized for di-electron spectroscopy and one for
di-muon spectroscopy. Open charm measurements will presumably rely on
the CBM di-electron setup shown in Figure \ref{GlobalCBM} (left). This
setup is formed by a Micro Vertex Detector (MVD) and a Silicon
Tracking System (STS), which operate in a 1 Tm magnetic
field. Electron identification is provided by a RICH at lower particle
energies and by a set of Transitions Radiation Detectors (TDR) at
higher energies. A Time-Of-Flight (TOF) system aims for the
identification of hadrons with low and medium energies. The setup is
completed by an electro-magnetic calorimeter, which allows the
measurement of direct photons and by a forward hadronic calorimeter
(not shown), which measures the energy of spectators of the nuclear
collision.

Figure \ref{GlobalCBM} (right) shows a zoom into the region of MVD and
STS. Both systems are formed by planar detectors. The outer acceptance
angle of both detector system (and all CBM) is given with
\mbox{$\vartheta = 25^{\circ}$} with respect to the beam axis. An
inner opening of the detector stations, which is to allow for a
passage of the beam pipe, limits the inner acceptance of the
experiment to \mbox{$\vartheta = 2.5^{\circ}$}.  The MVD will operate
in the moderate vacuum of the beam pipe which might be separated from
the vacuum of the SIS300 synchrotron by differential pumping or a thin
foil located afar from the experiment in the beam pipe. The aim of
this concept is to avoid unwanted multiple scattering of the particle
tracks in a vacuum window located between the target and the MVD. In
our concept, this vacuum window will be located between the MVD and
the STS. The latter operates in the cooled atmosphere required for
avoiding unwanted radiation damage effects like intense leakage
currents or reverse annealing in the silicon strips.

The details of both, the MVD and the STS, are still being
optimized. Presumably the first detector station of the MVD will be
located \mbox{5 - 10 cm} downstream the target. It will use ultra thin
and highly granular silicon pixel detectors while the STS relies on
radiation hard, double sided strip detectors. It is still debated if
some intermediate layers of very fast silicon pixel detectors might be
beneficial for tracking.

CBM is designed as free running system using self-triggered detectors
and high level event selection. The trigger concept for open charm
measurements aims to pick up the zero suppressed data stream provided
by the detectors and to reconstruct the event by performing tracking
in the MVD and STS in real time. A scan for displaced decay vertexes
is intended to allow for a selection of interesting events. The
details of this tracking concept are still under debate. The task is
complicated by the presence of displaced decay vertexes from the
decays of strange particles.

\section{Fundamental considerations on the requirements on the CBM-MVD}
\label{RequirementsMVD}
\begin{figure}[tb]
  \begin{minipage}[c]{8cm}
    \begin{center}~\\\hspace{1.5cm}\\
      \includegraphics[width=8cm]{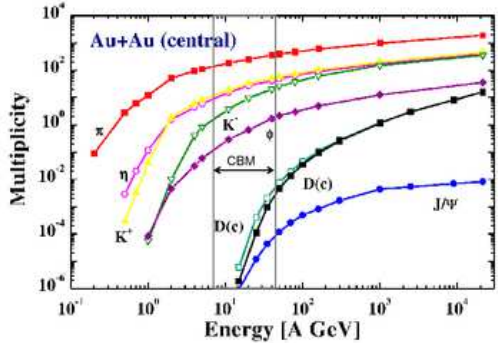}
    \end{center}
  \end{minipage}
  \begin{minipage}  [c]{8cm}
    \caption[Multiplicities of strange and charmed particles being
      emitted from a Au+Au collision.]  {\it \small The average number
      of mesons produced per central Au+Au collision (multiplicity) as
      a function of the incident beam energy. The calculation was
      performed with the HSD transport code. No in-medium mass
      modification was taken into account. As open charm (D) is
      produced close to the kinematic threshold in the SIS-300 energy
      range (10 - 40 AGeV), the production multiplicity is small and
      varies strongly as function of the beam energy. Figure from
      \cite{OpenCharmProduction}.
      \label{CharmMultiplicity}
    }
  \end{minipage} 
\end{figure}

\subsection{Beam time and collision rates}

The requirements on the CBM-MVD are derived from two elementary needs,
which are the production and the reconstruction of open charm. At
SIS-300 energies, open charm production occurs close to the kinematic
threshold. The multiplicities are therefore low and, due to a lack of
experimental knowledge on the elementary cross sections, difficult to
quantify from theoretical models. According to the predictions
\cite{OpenCharmProduction} shown in figure \ref{CharmMultiplicity}, we
consider a production multiplicity of roughly $10^{-5} - 10^{-3}$ open
charm particles of each species ($\rm D^{\pm},~D^0,~\Lambda_C$) per
central Au+Au collision.  Accounting for the branching ratio of the
hadronic decay channels and assuming a reconstruction efficiency of
few percent, one estimates that one may reconstruct roughly $10^{-8} -
10^{-6}$ open charm particles of each species per central
collision. We aim to measure $\sim 10^{10}$ to $10^{12}$ central
collisions to reconstruct roughly $10^4$ open charm particles of each
flavor per year.

It is planned that within one year, CBM will have \mbox{5 $\cdot 10^6$
  s} \mbox{(two months)} beam on target. One requires therefore a
minimum collision rate of $2 \cdot 10^3$ to $2 \cdot 10^5$ central
collisions per second, which corresponds to $2 \cdot 10^4$ to $2 \cdot
10^6$ collisions integrated over all impact parameters. Using an 1\%
interaction target, this rate is in reach of the very high intensity
values of the SIS-300 synchrotron. However, it introduces strong
constraints on the time resolution and the radiation hardness of the
detector system.

\subsection{Radiation doses}

In order to estimate the requirements on the radiation hardness of the
MVD, the expected radiation doses for a vertex detector station
located $z=5~\rm cm$ and $z=10~\rm cm$ downstream the target were
simulated with GEANT-3~\cite{Geant3} + GCALOR~\cite{GCalor}. The results
of this exploratory study were confirmed with a comparative study
using FLUKA~\cite{FLUKA}. Both studies simulated radiation doses
obtained in 25 AGeV Au+Au collisions with random impact parameter,
which were generated with UrQMD~\cite{URQMD1} .  The non ionizing
energy loss of the particles penetrating the detector stations were
set according to the tables presented in~\cite{NIEL-tafel}.  The
numbers were normalized assuming a beam intensity of $10^9$ ions per
second and a $1\%$ target generating $10^7$ collisions per second. The
integrated yearly operation time of CBM was set to \mbox{$5 \cdot
  10^6~\rm s$} beam on target.

\begin{figure}[tb]
  \begin{minipage}[c]{8cm}
    \includegraphics[width=7cm]{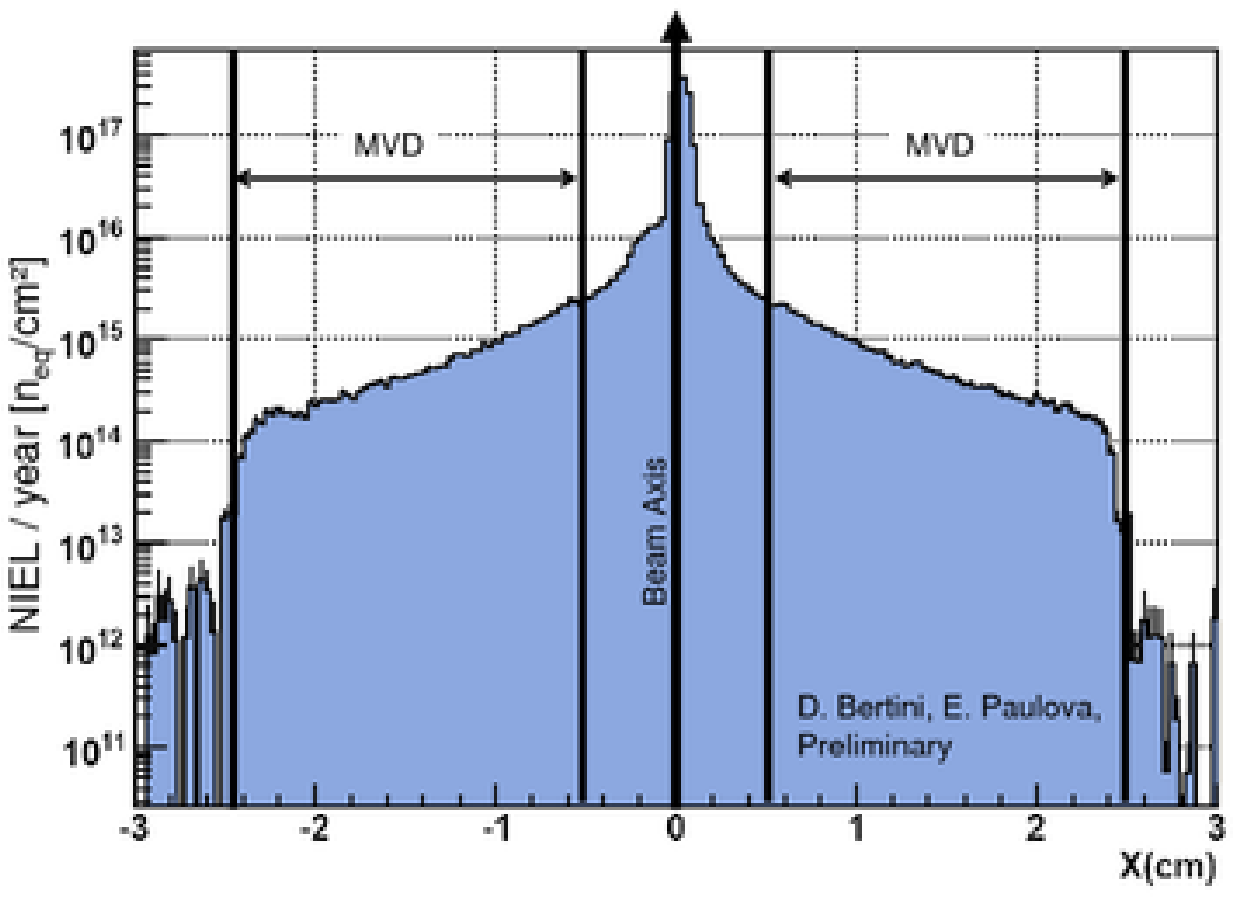}
  \end{minipage}
  \begin{minipage}[c]{7.5cm}    
    \includegraphics[angle=90,width=7cm]{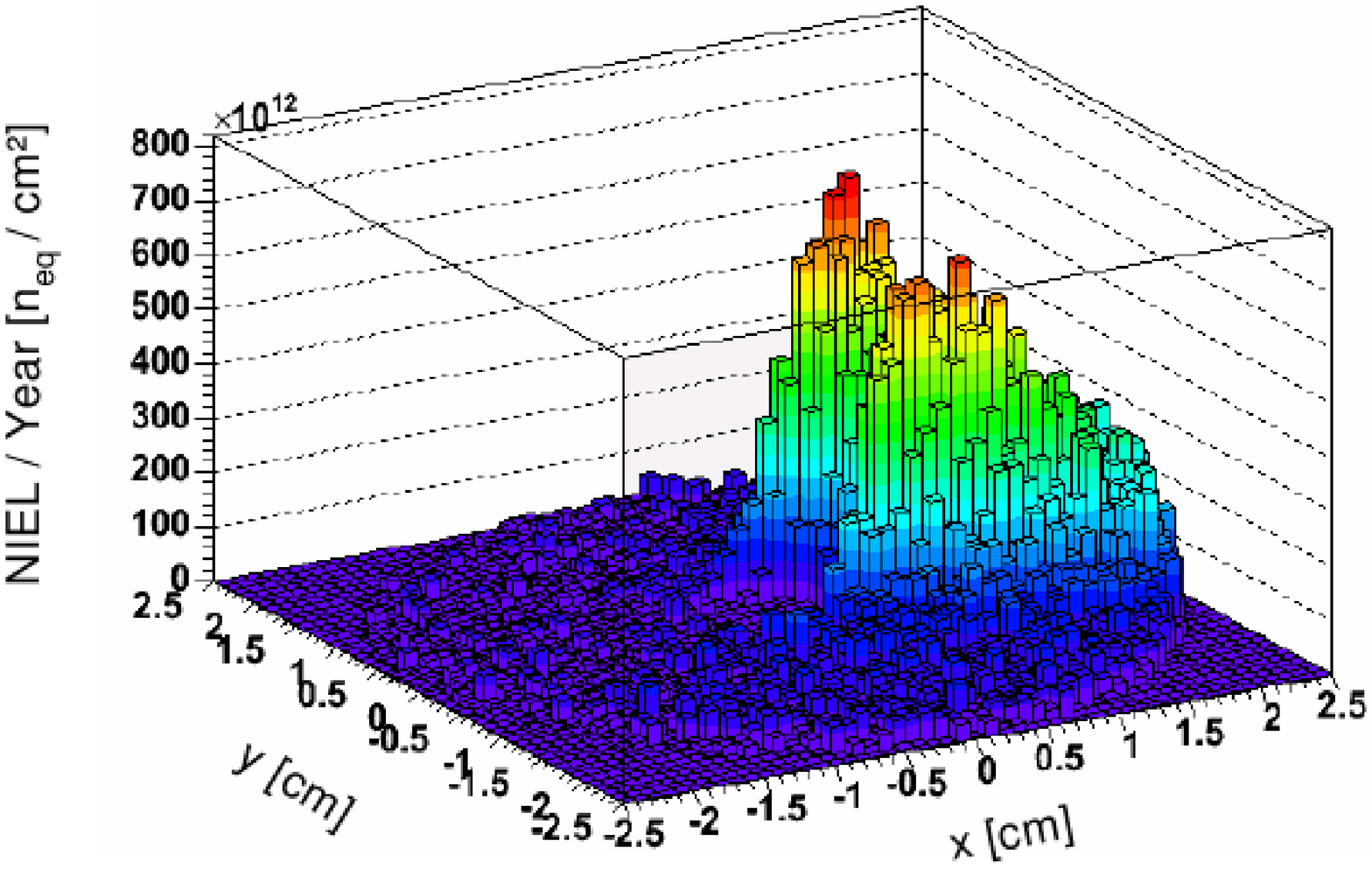}
  \end{minipage}  
  \caption[] {\\ \it {\bf Left:} Radiation dose for a vertex detector
    station located 5 cm from the target. The beam axis and the region
    covered by the MVD station according to the CBM standard geometry
    is shown.\\ {\bf Right:} The additional contribution caused by
    $\delta$-electrons (see text). \label{fig:Radiation-dose}}
\end{figure}

The preliminary results of the studies are shown in Figure
\ref{fig:Radiation-dose} (left) for a detector station located at
$z=5~\rm cm$. One observes that the radiation dose on the station is
largest close to the beam axis. At the border of the beam hole of the
detector station, it is up to \mbox{$\sim 2 \cdot 10^{15}~{\rm
    n_{eq}/cm^2}$} per year. Over the $2~\rm cm$ covered by the
station, the radiation dose drops by one order of magnitude.

We noticed that a sizable amount of $\delta$-electrons are knocked out
from our target by the primary beam. Despite a part of these electrons
are deflected by the 1 Tm dipole field of the tracking magnet, they
contribute substantially to the non-ionizing radiation. This
contribution was simulated with GEANT-3 + GCALOR by shooting gold ions
through the gold target of the experiment. The yield and spectrum of
$\delta$-electrons generated by this engine were checked against
\cite{deltaElectronEquation}. In the electron momentum region of
interest between $p=10~\rm MeV$ and $p=100~\rm MeV$, the simulation
results were found to exceed the theoretical prediction by few 10\%,
what is neglected in the following. Again the damage factor of the
electrons was set according to \cite{NIEL-tafel}. Despite their small
damage factor in the energy region of interest (\mbox{$0.05 - 0.09
  ~{\rm n_{eq}/cm^2}$}), the electrons provide an additional radiation
dose of up to \mbox{ $\sim 0.8 \cdot 10^{15}~{\rm n_{eq}/cm^2}$} in
the hottest areas of the detector. As illustrated in Figure
\ref{fig:Radiation-dose} (right), the magnetic bending of the electron
tracks distributes this radiation dose in a very asymmetrical way over
the surface of the MVD station.

A station located at $z=10~\rm cm$ would still receive a non-ionizing
dose of \mbox{$\sim 1 \cdot 10^{15}~{\rm n_{eq}/cm^2}$} (plus
\mbox{$\sim 0.15 \cdot 10^{15}~{\rm n_{eq}/cm^2}$} from
$\delta$-electrons).

The $\delta$-electrons dominate by far the ionizing radiation dose in
the vertex detector. Assuming that all charged particles penetrating
the detector station are approximately minimum ionizing, the hottest
region of a detector station located at $z=5~\rm cm$ may accumulate
\mbox{$\sim 340~\rm Mrad$} per year. Additional magnetic deflection of
$\delta$ electrons reduces this dose to \mbox{$\sim 80~\rm Mrad$} for
a station located at $z=10~\rm cm$.

\subsection{Vertex resolution}
\label{ChapterFundamentalConsiderations}
Our main selection criterion for identifying open charm particles will
be to separate their secondary decay vertex from the primary
vertex. To do so, the vertex detector has to extrapolate the
trajectories (i.e. of particle pairs) back to their intersection
point, which is typically equal to the primary vertex. If both
particles are decay products of an open charm particle, their
intersection point is away from this primary vertex as the open charm
particle traveled a certain distance before decaying.

In the previous section, we assumed a reconstruction efficiency of few
percent for open charm mesons. Following this requirement, one can
obtain a first impression about the necessary secondary vertex
resolution of the detector. To do so we assume in a simplistic
estimate, that CBM can trigger on central events. We consider for
reasons of simplicity that the separation of the secondary vertex from
the primary vertex should be the sole cut within the analysis. The
indicated intersection points of particle pairs generated in the
primary vertex should be distributed around this vertex according to a
Gaussian distribution. Moreover, the velocity of the open charm
particles should be equal to the center of mass velocity of the
collision system.  Be $z=0~\rm cm$ the position of the target and thus
of the primary vertex. According to the decay law, the distribution of
the decay length of open charm is given with:
\begin{equation}
n_{_{\rm S}}(z) = C_0 \cdot \exp \left ( - \frac{z}{\beta \gamma \cdot c \tau} \right )
\label{GleichungZerfallD0}
\end{equation}

\begin{figure}[tb]
  \begin{minipage}[c]{8cm}
    \includegraphics*[width=7cm]{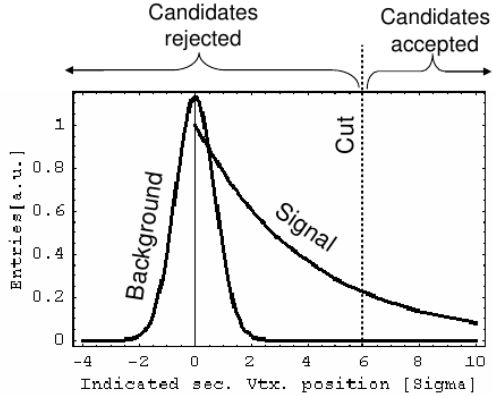}
  \end{minipage}
  \begin{minipage}[c]{8cm}
    \caption[Significance Estimate] {\it Illustration of
      $n_{_{D^0}}(z)$ (signal) and the corresponding background as a
      function of the reconstructed position of the displaced
      secondary vertex ($z$). This position is given in units of the
      secondary vertex resolution $\sigma_{_{SvZ}}$.  See text.  }
   \label{SignificanceEstimate}
  \end{minipage}   
\end{figure}

In this equation, which is illustrated in figure
\ref{SignificanceEstimate}, $z$ stands for the distance between the
reconstructed secondary decay vertex of, for example, an $\rm D^0$ and
the primary vertex. $C_0$ is a normalization factor, which will cancel
out in the following. The Lorentz boost of the center-of-mass for a
beam energy of \mbox{$25~\rm AGeV$} is \mbox{$\gamma = 3.8$}, which
allows us to set the velocity of the particles to $\beta \approx 1$ in
the following.

In order to obtain a good reconstruction efficiency of 5\%, we chose a
parameter $z_0$ such that a fraction of 5\% of all open charm
particles are within the selection criteria $z>z_0$. This is fulfilled
if:

\begin{equation}
F_{_{\rm S}}(z_0)= \frac{A \cdot N_{\rm S}}{N_{\rm All}} =
\frac{A\cdot \int\limits_{\rm z_0}^{\infty} n_{_{\rm S}}(z)
  ~dz}{\int\limits_{0}^{\infty} n_{_{\rm S}}(z) ~dz}> 5\%
\label{equationFractionSvZ}
\end{equation}
In the equation, $N_{\rm S}$ stands for the selected open charm
particles, $N_{\rm All}$ stands for the total number of created open
charm particles and $A \approx 0.35$ for the geometrical acceptance of
CBM. The condition is fulfilled if:
\begin{equation}
z_0 \leq 2 \cdot \gamma \cdot c \tau
\label{eq:SignalD0}
\end{equation}
This result sets a first constraint on our cut.

In a next step we assume, that we want to reach a purity of the signal
of $S/B=1$. As explained above, $S=10^4$ open charm particles have to
be reconstructed and that therefore $B=10^4$ background particles can
be tolerated within the run. Assume that this run measured $10^{10}$
central Au+Au collisions with 25 AGeV. As each collision generates
roughly 400 negatively charged and 500 positively charged particles,
$2\cdot10^5$ combinations per collision are to be considered if two
body decays are analyzed.

The total background of a single run is therefore formed by $N_{\rm
  BC}= 2 \cdot 10^{15}$ background candidates among which $N_{\rm
  ABC}=A \cdot N_{\rm BC}=7 \cdot 10^{14}$ are geometrically
accepted. The indicated origin of those pairs follows a Gaussian
distribution with a width, which is equal to the secondary vertex
resolution of the vertex detector $\sigma_{\rm z}$. In order to reach
a background of $B=10^4$ only, the selection has to reduce it by a
factor of $r \leq 7 \cdot 10^{-10}$. The relative number of entries
above a certain cut $z_0$ in a Gaussian is given by:
\begin{equation}
r = \frac{1}{2} \cdot \left (1- {\rm Erf} \left [\frac{ z_0
  }{\sigma_{\rm z} \cdot \sqrt{2}} \right ] \right)
\end{equation}
The condition $r\leq 7 \cdot 10^{-10}$ is fulfilled if:
\begin{equation}
z_0 \geq  6.1 ~\sigma_{\rm z}
\label{eq:BgGauss}
\end{equation} 
Combining equation \ref{eq:BgGauss} and equation \ref{eq:SignalD0}, one concludes that
\begin{equation}
\sigma_{\rm z} = 0.3 \cdot \gamma \cdot c \tau
\end{equation}
In order to reconstruct our most challenging observable, the
$\Lambda_C$, reasonably well, we would need a secondary vertex
resolution of:
\begin{equation}
\sigma_{\rm z} = 0.3 \cdot \gamma \cdot c \tau(\Lambda_C)=0.3 \cdot 3.8 \cdot 59.9 {~\rm \mu m} = 70{\rm ~\mu m}.
\end{equation}
This value gives a first estimate for the requirements of CBM in terms
of secondary vertex resolution.

It should be mentioned that the simplistic calculation shown here
comes with several optimistic assumptions. In particular it neglects
the presence of secondary decay vertexes originating from the decay of
strange particles. Consequently, the detector performances suggested
by the calculation are too optimist. Nevertheless, the approach
provides a reasonable lower limit for the requirements of the detector
system. Simulation results allowing for an estimate of the physics
potential of the full detector system will be discussed in section
\ref{SectionSimulation}.

\subsection{Spatial resolution and material budget}

Systematic simulations and with CBMROOT \cite{FairRoot} were performed
in order to translate the requirements on the secondary vertex
resolution into requirements of the detector. The secondary vertex
resolution of the MVD was studied as function of the material budget
and spatial resolution of the detector stations. Tracks having a
momentum of \mbox{$p\leq \rm 1~GeV/c$} were ignored as they are
typically rejected within our data analysis for searching open charm
signatures. The spacial resolution of the MVD stations was simulated
with Gaussian smearing of the hit position.

\begin{figure}[tb]
  \begin{minipage}[c]{8cm}
    \includegraphics[width=8cm]{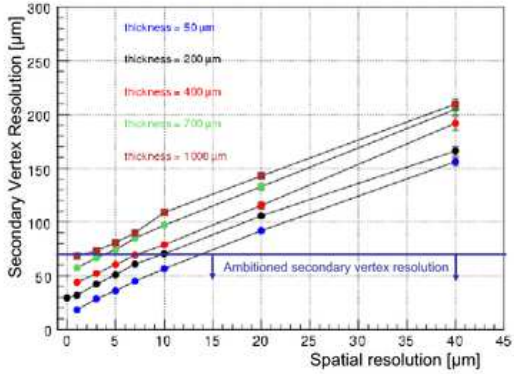}
  \end{minipage}
  \begin{minipage}[c]{8cm}
    \caption[Significance Estimate] {\it The secondary vertex
      resolution of the CBM-MVD as function of the material budget
      (expressed in $\rm \mu m$ silicon) and the spatial resolution of
      the detector. Note that \mbox{$1000~{\rm \mu m ~ Si}\approx 1\%~
        X_0$}. Only tracks with \mbox{$p>1~\rm GeV/c$} were accounted
      for.  }
    \label{SvZResolution}
  \end{minipage}   
\end{figure}

The results of the simulation for a first detector station located at
\mbox{$z=5~\rm cm$} are shown in figure \ref{SvZResolution}. One
observes that the secondary vertex resolution of the MVD increases
roughly linearly with the spatial resolution of the detector
stations. A spatial resolution of \mbox{$\sigma \lesssim 10 \mu \rm
  m$} (in both dimensions!) in combination with an ideally thin
detector seems mandatory to fulfill the requirements of CBM. The hard
maximum for the material budget would be \mbox{$x \lesssim 1\%~X_0$}
for detectors with "ideal" spatial resolution. A realistic combination
of both parameters would be a spatial resolution of \mbox{$\sigma
  \approx 5~\mu m$} and a material budget of of the detector stations
of \mbox{few 0.1\% $X_0$} (corresponding to silicon with a few $100
\rm ~\mu m$ thickness).

\section{Technological constraints for the design of the MVD}
\label{ChapterTechnologicalConstraints}
\subsection{Technological options}

\begin{table} [t]
\centering
\small
\begin{tabular}	{||c||c||c|c|c||}
		\hline
		\hline
{\bf } & {\bf Required$^{(1)}$} & {\bf Hybrid } & {\bf CCD}& {\bf MAPS } \\
		\hline
		\hline
{Spat. resol. [$\rm \mu m$]} & {$\lesssim 5$} & {$\sim 30 ~^{(2)}$} & {$\sim 5~^{(2)} $} & {$\lesssim 3$} \\
		\hline
{Mat. budget [$\rm X_0$]} & {few $0.1\%$}& {$\sim 2\% ~^{(3)}$} & {$\sim 0.1\%~^{(4)}$} & {$\sim 0.05\%~^{(4)}$} \\
		\hline
{Rad. hardn. [$\rm n_{eq}/cm^2$]} & {$\rm few ~ 10^{15}$/year } & {$\sim 10^{15}$} & {$\sim 10^{10}$} & {$ \gtrsim 10^{13}$}\\
		\hline
{Time resolution} & {$\lesssim 100~\rm ns$} & {$25~\rm ns $} & {$\sim 50~\rm \mu s~^{(5)}$} &{$\sim 20~\rm \mu s$}\\
		
		\hline
		\hline

		\end{tabular}

\caption[Performances of different pixel detectors]{ \it Performances of
  different pixel detectors compared to the requirements for open
  charm meson reconstruction with full collision rate at CBM. The data
  on hybrid pixel detectors and CCDs was collected from
  \cite{CMS-Pixels},\cite {Atlas-Pixels}, \cite{CCD-Detectors} and
  \cite{CCD-Irrad}. \\ {\footnotesize {\bf Remarks:} $^{(1)}$: For
    operating the MVD at the SIS-300 top luminosity.  $^{(2)}$:
    Derived from the typical pixel pitch assuming digital readout.
    $^{(3)}$: ATLAS pixel module.  $^{(4)}$: Sensor thickness.
    $^{(5)}$: Design goal for the International Linear Collider.  } }
\label{VergleichPixelDetektoren}				
\end{table}

The requirements derived so far are listed in Table
\ref{VergleichPixelDetektoren}. They are valid for operating the Micro
Vertex Detector at the CBM top luminosity, which is mandatory for open
charm measurements closest to production threshold. Due to the higher
production multiplicities, less stringent performances in terms of
time resolution and radiation hardness are sufficient in order to do
open charm physics at the higher beam energies of CBM. The table also
provides information about the typical performances of established
pixel detector systems like hybrid pixels and CCDs. Obviously, both
detector concepts do not match the challenging requirements as the
very radiation hard and fast hybrid pixel detectors do not reach the
necessary spatial resolution and show a too large material budget. The
very light and granular CCDs miss the requirements in terms of
radiation hardness by many orders of magnitude and would therefore
fail within minutes due to radiation damage.

As the existing pixel detectors do not match the requirements of the
CBM experiment, we searched for alternative technologies. The most
promising candidates were found among the pixel detector systems being
developed for the International Linear Collider (ILC), as this
experiment has similar requirements in terms of granularity and
material. Among the existing concepts, we identified
DEPFETs~\cite{depfetPaper} and CMOS Monolithic Active Pixel Sensors
(MAPS)~\cite{MAPSProposal,FirstMAPSTest,MAPS-TestPaper,MIMOSA4Paper}
as most interesting options. We chose MAPS as non-exclusive guide line
technology as their development had further progressed and as
operating DEPFETs in our very inhomogeneous ionizing radiation fields
appears challenging\footnote {DEPFET detectors demonstrated a
  competitive radiation tolerance against \mbox{$\gtrsim 1 ~\rm Mrad$}
  \cite{DepfetPaperStrahlung}. However, operating the irradiated
  detector requires to adapt the gate voltage of the DEPFET in order
  to compensate its radiation induced threshold voltage shift. In
  vertex detectors with a collider geometry, one expects in first
  order a gradient in radiation doses in one dimension. This allows
  setting a common compensation voltage for all pixels of a line,
  which requires only moderately adapted steering chips. In contrast,
  the specific case of the CBM-MVD with its strong gradients in
  radiation doses in \emph{both} chip dimensions might require to set
  a compensation voltage for each individual pixel. The latter would
  call for a steering logic of challenging complexity.}.

Comparing the performances of MAPS with the requirements of the
CBM-MVD, one observes that the sensors cannot cannot match all
requirements. However, they combine the strong points of CCD-detectors
with a by three orders of magnitude higher radiation tolerance. MAPS
were therefore considered to provide the best technological compromise
available today. Therefore, we decided to work out, which part of the
CBM physics program one might cover with this technology.

\subsection{Features of Monolithic Active Pixel Sensors}

CMOS Monolithic Active Pixel Sensors for charged particle tracking
were initially developed for the vertex detector of the International
Linear Collider by the IPHC Strasbourg. They were derived from sensors
used for optical imaging.  A single point resolution of \mbox{$1 - 2
  ~\mu$m} and a detection efficiency close to 100\% were routinely
observed in beam tests at the CERN-SPS with various MAPS designs
featuring up to 10$^6$ pixels on active areas as large as
\mbox{$4~$cm$^2$}.

Radiation hardness studies with different MAPS-prototypes
\cite{neutronPaper1,neutronPaper2} showed that radiation significantly
increases the leakage currents of the collection diodes of their
pixels. A moderate cooling of the sensor allows keeping these leakage
currents at a level where the corresponding shot noise remains
marginal. Rather a deterioration in the Charge Collection Efficiency
(CCE), translating into a decrease of the S/N of the sensors, was
identified as the key limitation for the radiation hardness of MAPS. A
possible explanation is that an increased bulk damage in the epitaxial
layer strongly reduces the lifetime of the diffusing charge carriers.
It is considered that therefore the lifetime of the signal electrons
in the sensor drops below the value required for collecting them by
thermal diffusion. Shortening the diffusion paths by choosing a
smaller pixel pitch alleviates this effect and allowed recently
reaching a radiation tolerance against \mbox{$\gtrsim 10^{13}~$n$_{\rm
    eq}/$cm$^2$} \cite{annualReport}.

\begin{figure}[t]
  \begin{center}   
    \begin{minipage}[c]{8cm}
      \includegraphics[width=8cm]{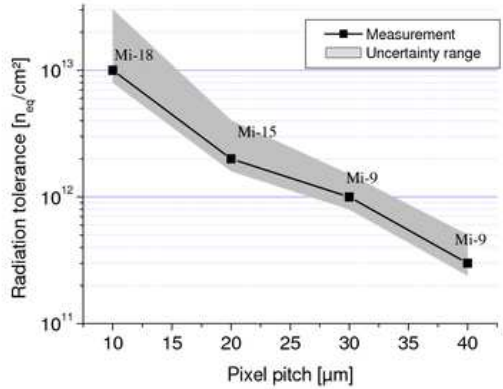}
    \end{minipage}
    \begin{minipage}[c]{8cm}
      \caption[] {\it The measured tolerance of MAPS prototypes against
        non-ionizing radiation as function of the pixel pitch.
 \label{radiationDoseFigure}}
  \end{minipage}
\end{center}   
\end{figure}

Remarkably the correlation of the radiation tolerance of the sensors
and the length of the diffusion paths turns into a correlation between
this radiation tolerance and the pixel pitch of the sensors. This is
illustrated in Figure \ref{radiationDoseFigure} which shows the
measured radiation hardness of MAPS-prototypes with different pixel
pitch manufactured in the AMS 0.35 $\rm \mu m$ Opto process.

An empirical fit of the measured data suggests that the radiation
hardness of MAPS against non-ionizing radiation scales roughly
according to:
\begin{equation}
T_{non-io} \approx 1.89\times 10^{15}\rm ~ n_{ \rm eq}/cm^2 \cdot \left(\frac{P}{\mu m}\right)^{-2.27}
\label{radhardEquation}
\end{equation}
In this equation, $T_{non-io}$ stands for the tolerance of the sensors
against non-ionizing doses and $P$ for the pixel pitch\footnote{Note
  that the radiation hardness of MAPS depends on some parameters,
  which are specific to the CMOS process used for their production
  (namely the thickness of the epitaxial layer). Equation
  \ref{radhardEquation} does therefore not claim a general validity.}.

The radiation tolerance of MAPS against ionizing doses is currently given with \mbox{$1~\rm Mrad$} independently of the pixel pitch \cite{Doktorarbeit}.

\subsection{Sensor geometry and time resolution}
\label{ChapterTimeResolution}

MAPS are monolithic detectors which integrate the readout electronics
and the sensor on the same, back thinned CMOS chip. Each pixel hosts a
preamplifier located on top of the sensitive volume of the sensor,
which is the epitaxial layer of the chip. A P-Well implantation
hosting the transistors of the amplifier is used to isolate those
transistors electrically from the sensitive layer. As any N-Well
implantation other than one of the N-Well/p-epi collection diode would
generate a parasitic charge collection, no PMOS transistors can be
used in the sensitive surface. Because of this constraint, all logics
requiring those transistors (for example discriminators) have to be
placed at a separate surface outside the pixel matrix.

\begin{figure}[t]
  \begin{center}   
    \begin{minipage}[c]{10cm}
      \includegraphics[width=9cm]{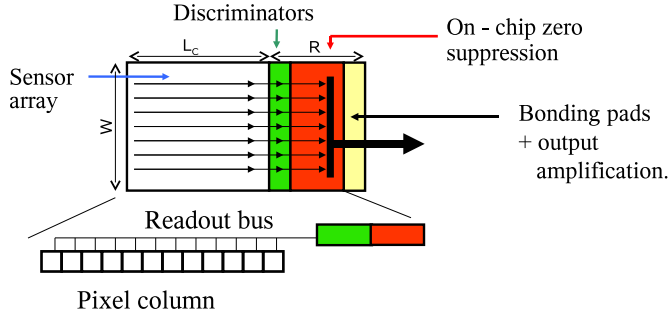}
    \end{minipage}
    \begin{minipage}[c]{6cm}
      \caption[] {\it The concept of MAPS with massive column parallel
        readout. The signal of the pixels of each column is
        multiplexed on one readout bus and shipped towards a
        discriminator and data sparsification circuits located aside
        the pixel matrix. See text.
        \label{MimosaConceptFigure}}
    \end{minipage}
  \end{center} 
\end{figure}

The readout of fast MAPS is therefore done in the massive column
parallel way, which is illustrated in figure
\ref{MimosaConceptFigure}. The signal of the pixels of a column (or
line) are multiplexed on one common readout bus and shipped to a
discriminator being located aside the pixel matrix. The data of the
discriminators is received by a digital data sparsification circuit
which is to execute zero suppression. The compressed hit information
is written out towards the DAQ of the experiment. The feasibility of
this concept has meanwhile been demonstrated experimentally by
building and testing the sensor matrix including the discriminators
and, on a separated chip, the data sparsification circuits
\cite{Mimosa22-Paper}.

The concept of the chip readout introduces several constraints for the
global design of the CBM-MVD. The most important one is a constraint
in terms of time resolution and readout speed, which is caused by the
limited bandwidth of the column readout bus. It is expected today,
that this bus may allow for \mbox{$f \approx 10^7$} readout processes
per second\footnote{A readout with $f = 6~\rm MHz$ is demonstrated
  with sensors manufactured in the AMS 0.35 $\mu$m Opto process.}. The
readout time $t_{int}$ of a column with $N_P$ pixels is therefore
given with:
\begin{equation}
t_{int}= \frac{N_{P}}{f} 
\label{Eq:ReadoutSpeed}
\end{equation}
Equation \ref{Eq:ReadoutSpeed} is of particular importance as it
connects the time resolution of the pixel matrix with its geometrical
surface. Knowing that each pixel has a pitch $P$, one can derive the
maximum length $L_C$ of the pixel matrix in one dimension. This is
given with:
\begin{equation}
L_C= N_{P} \cdot P = t_{int} \cdot f \cdot P
\end{equation}
The width $W$ of a pixel matrix is constrained by the size of a reticle, which is between 20 and \mbox{30 mm} depending on the CMOS process used for the production of the sensor.

The discriminators and data sparsification circuits are located aside
the pixel matrix. The surface required for this logic in the direction
of $L_C$ is expected to be roughly \mbox{$R=1-3 \rm mm$}. The readout
electronics covers therefore a surface of $R \cdot W$. This surface is
passive and needs to be covered by the pixel matrix of a second chip
in order to reach a 100\% fill factor in the MVD. Assuming that for
reasons of material budget reduction, not more than two layers of
silicon are acceptable, this pixel matrix must have the same surface
than the surface covered by the readout electronics. From this
constraint one derives the theoretical time resolution of MAPS, which
is given with:
\begin{equation}
t_{int}=\frac{R}{P \cdot f}
\label{Eq:ReadoutSpeed2}
\end{equation}
Plausible values for the pixel pitch are between $P=10~\rm \mu m$ and
$P=30~\rm \mu m$. From this, one can derive the approximative time
resolution of MAPS, which is given with $t_{int} \approx 10 ~ \rm \mu
s$.  As this time resolution is by up to a factor of 100 longer than
the mean time between two collisions at the CBM maximum collision
rate, one expects a pile-up of nuclear collisions in the
MVD. Disentangling this pile-up is one of the major challenges for the
tracking algorithms of CBM. It is considered to start the tracking at
the most downstream STS detectors and to extrapolate identified tracks
toward the MVD. Due to the very good granularity of MAPS, we suppose
that a moderate pile-up of nuclear collisions will not translate into
an excessive detector occupancy. However, track densities and the
occupancy of the MVD is a crucial topic.

\subsection{Track densities}

In order to obtain a first estimate about the occupancy of the MVD, we
performed simulations using CBMROOT framework and the GEANT-3+GCALOR
engine. As already done for the radiation dose simulations, we
accounted for two sources of particles, which are the the primary
particles generated in the nuclear collisions and the
$\delta$-electrons, which are knocked out from the target by beam
ions.  The latter cannot be ignored as, unlike the faster hybrid pixel
and strip detectors, MAPS pile up all $\delta$-electrons produced
between the nuclear collisions. Assuming a 1\% interaction target, the
particles of a primary collision are complemented by the
$\delta$-electrons produced by 100 heavy ions passing target. This
makes those electrons a crucial contributor to the occupancy of the
pixels.

We simulated the relevant track density under two assumptions. In the
first simulation we assumed that the beam intensity is adapted in such
a way to the ability of the detector, that the detector can
distinguish the individual collisions. In this case, one will try to
select central collisions and the occupancy of the detector is
determined by the pileup of a central collision and the
$\delta$-electrons from 100 ions passing the target. In the second
case, the detector will face higher beam intensities and therefore
have to handle tracks originating from several nuclear collisions. As
a selection of central collisions is not further feasible, this
scenario is best described by merging $\delta$-electrons with a
central collision and further nuclear collisions with random impact
parameter.
 
\begin{figure}[tb]
  \begin{center}   
    \begin{minipage}[c]{7cm}
      \includegraphics[width=7cm]{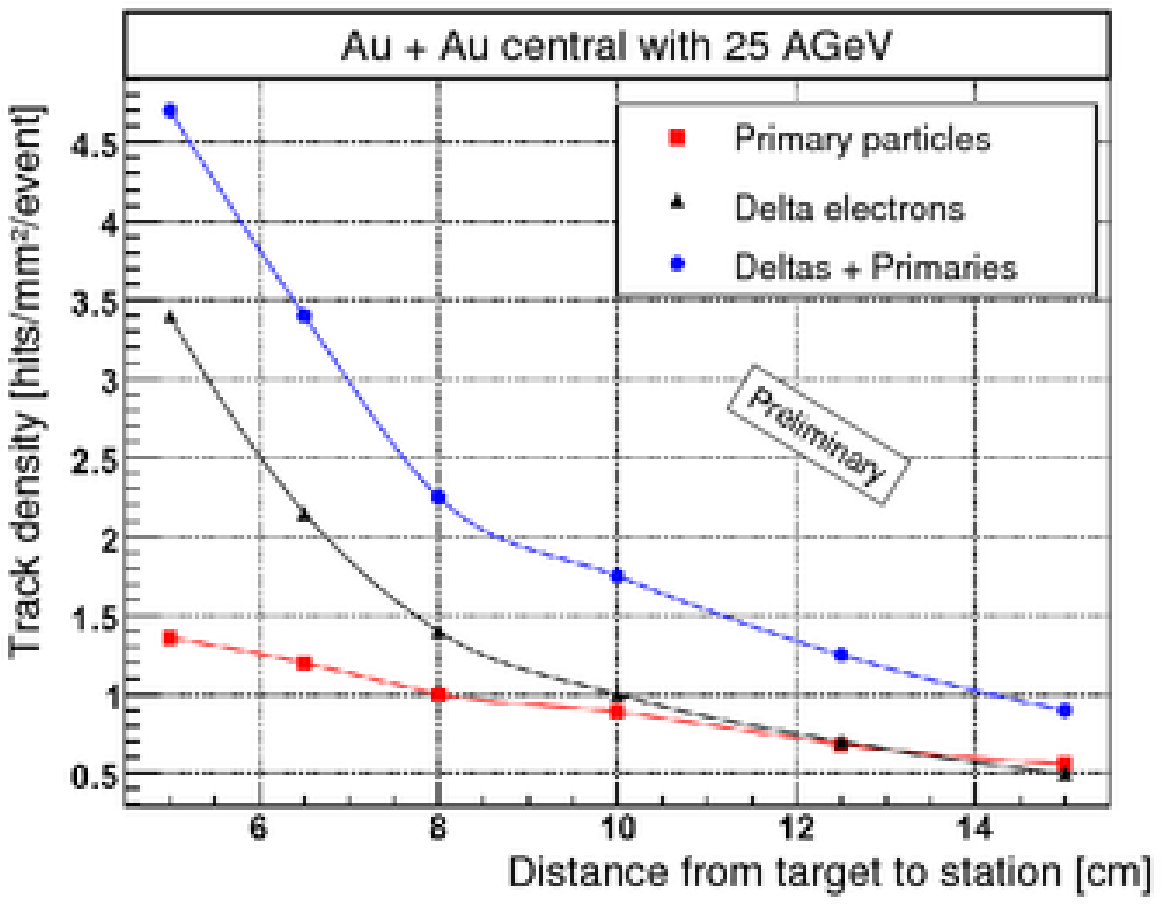}
    \end{minipage}
    \begin{minipage}[c]{7cm}
      \includegraphics[width=7.2cm]{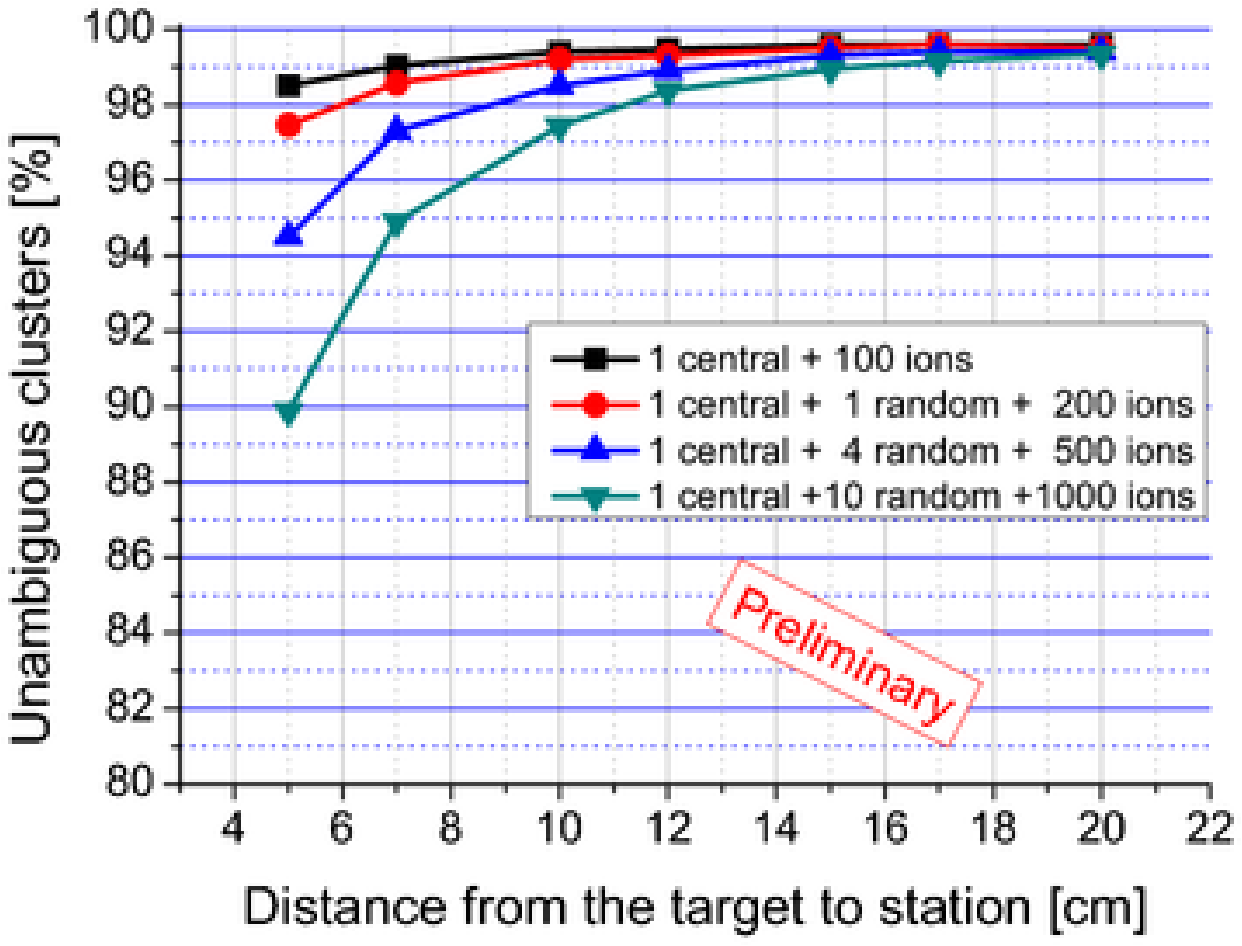}
    \end{minipage}
    \caption[] {\it \\{\bf Left:} The peak occupancy of a MAPS based MVD
      as function of the position of this station for central
      collisions.As the final design of the MVD is not yet fixed,
      several potential positions for the detector stations were
      studied.\\ {\bf Right:} Fraction of unambiguous clusters, which
      were well separated from clusters of neighboring tracks with
      respect to all clusters. The simulated events were composed by a
      pileup of a central collision, some collisions with random
      impact parameter and the delta electrons generated from 100 ions
      per collision. A pixel pitch of \mbox{$P=10~\rm \mu m$} was
      assumed.
      \label{Occupancy}}
  \end{center} 
\end{figure}

The preliminary simulation results for the track densities generated
by central collisions are displayed in figure \ref{Occupancy}
(left). The figure shows the peak track density of MVD stations as
function of the position of those stations.  It should be mentioned
that this track density shown is reached only at the small fraction of
the detector, which is most intensely bombarded with
$\delta$-electrons.

The results show that a station located at $z=5~\rm cm$ will face
track densities of up to \mbox{5 $\rm hits$} per $\rm mm^2$ and
collision already without pile-up. Knowing average number of firing
pixels in a cluster of a MAPS detector is $\sim 5$ and assuming a
small pixel pitch of \mbox{$P=10~\rm \mu m$}, this translates into an
acceptable occupancy of \mbox{$\sim 0.25\%$}. A pile-up of 10 nuclear
events, which corresponds to a beam intensity of \mbox{$\sim 10^6$}
collisions per second, would increase this occupancy to a considerably
high value of \mbox{$\sim 2.5\%$}.

Our worry concerning this occupancy is that reconstructed tracks might
pick up a wrong hit of a merged cluster in the first detector
station. Doing so might generate a slight modification of the track
extrapolation, which turns into false indications of displaced decay
vertexes. Given that CBM intends to trigger on those vertexes, already
a modest amount of such cases might question the trigger concept.

The question whether the MVD shows a sufficient hit separation
performance to exclude this scenario is being addressed with the newly
developed MAPS digitizer. Among others, this digitizer simulates the
charge sharing among the pixels of a cluster, a functionality which
was calibrated with data collected from beam tests of MAPS with a
\mbox{120 GeV/c pion} beam at the CERN-SPS. Moreover, the the software
package contains a low level cluster finding algorithm as it might be
used in the future experiment. Both features allow a realistic
simulation of the merging of clusters due to too high
occupancies. First and preliminary results from an ongoing study
suggest that hit merging is a crucial topic if the MVD operates with
"big" pixels of \mbox{$30~\rm \mu m$} and a sizable pileup. However,
as shown in figure \ref{Occupancy} (right), a combination of a modest
pile-up of below 10 events and the small pixels (\mbox{$10 ~\mu m$})
required for good radiation hardness reduces this effect substantially
and more than 90\% of all clusters well separated. Intense simulation
work has been started in order to estimate the impact of the
remaining, merged clusters on our tracking, the trigger concept and
the signal reconstruction abilities of CBM.

\subsection{Estimated power dissipation and the basic equation of MAPS}

The power consumption of the MAPS pixels is strongly dominated by the
one of the readout logic. This is as, except for the brief readout
phase, the power consumption of a pixel is only $\lesssim \rm 1 pW$
while the end of column discriminators and data sparsification blocks
have to operate continuously to handle the incoming multiplexed data
stream. For a first estimate of the power consumption of MAPS, one may
thus state that the power dissipation of MAPS based detector scales
with the number of columns required for using a unit of surface of the
detector. To estimate this number, one has to know the surface of the
individual column $S_C$, which is given with:
\begin{equation}
S_C=N_P \cdot P^2
\end{equation}
Here, $N_P$ stands for the number of pixels in this column and $P$
represents the pitch of those pixels, which are assumed as squared.
We set $P_{col} (f)$ the power consumed by one end of column block and
obtain a power density of a vertex detector surface:
\begin{equation}
\rho_{Power}= \frac{P_{col}(f)}{N_P \cdot P^2}
\end{equation}
Note that equation \ref{Eq:ReadoutSpeed} correlates $N_P$ with the
readout speed of the detector $t_{int}$. Merging both equations one
obtains:
\begin{equation}
\rho_{Power}= \frac{P_{col}(f)}{t_{int} \cdot f \cdot P^2}
\label{EquationEnergieDichte} 
\end{equation}
Introducing moreover the dependence of the radiation hardness from the
pixel pitch (see equation \ref{radhardEquation}) into this equation,
one obtains:
\begin{equation}
\rho_{Power}=  \frac{0.035}{\rm m^2 }\cdot \frac{ P_{col}(f)  }
{ t_{int} \cdot f } \cdot  \left(\frac{T_{non-io}}{~\rm 
n_{eq}/cm^2} \right)^{0.88}
\label{MapsBasisGleichung}
\end{equation}
This equation links the most important parameter of the sensor
technology, which is the radiation hardness, the integration time and
the power consumption\footnote{Note that the restricted validity of
  equation \ref{radhardEquation} applies also to equation
  \ref{MapsBasisGleichung}.}. In case the pixel pitch is determined by
the needs for good single point resolution $\sigma$, one can set (as
suggested by the beam test results shown in \cite{MAPSResolution})
\begin{equation}
\sigma= \frac{P}{5}
\end{equation}
and one obtains:
\begin{equation}
\rho_{Power}= \frac{P_{col}(f)}{t_{int} \cdot f \cdot (5 \sigma)^2}
\label{MapsBasisGleichung2}
\end{equation}
The design of a MVD station has to provide the necessary cooling power
to evacuate this heat load under vacuum conditions.

\section{The design approach for the CBM-MVD}
\label{SectionLadderConcept}

\subsection{The design concept}

\begin{figure}[t]
  \begin{center}\hspace{2cm}  
    \includegraphics[width=12cm]{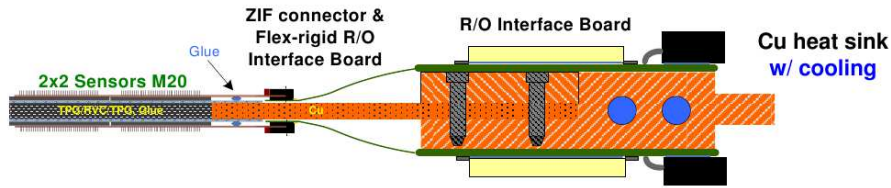}    
    \caption[] {\it The conceptual design of a ladder of the CBM-MVD
      detector. The ladder is formed by a layer of RVC-carbon
      foam. Two layers of the highly heat conductive TPG transport the
      heat produced by the sensors toward a heat sink outside the
      detector acceptance.  Flat band cables are used to bias the
      sensors and to transport the data to frontend boards located on
      top of the heat sinks.
      \label{LadderDesign}}
  \end{center}   
\end{figure}

In order to fulfill the requirements discussed in the previous
section, the design of our detector ladders follows the concept shown
in figure \ref{LadderDesign}. This figure displays a ladder, which is
formed by a mechanical support, silicon detectors and a heat sink,
which is located outside the acceptance of the experiment. The
mechanical support is composed from a sandwich of the ultra light
Reticulated Vitreous Carbon (RVC) foam and the highly heat conductive
Thermal Pyrolytic Graphite (TPG), which provides a very good heat
conductivity of \mbox{1700 W/m/K} at room temperature. While the RVC
delivers the necessary mechanical stability, the TPG transports the
heat produced by the sensors to the heat sink. The latter is cooled
with conventional liquid cooling and, as it is located outside the
detector acceptance, may contain a sizable amount of material.

The mechanical support hosts two layers of sensor chips. The sensors
of each layer are arranged to overlap the passive surface of the
sensors of the opposite layer, which allows to cover the planar
surface of a MVD station with an (almost) 100\% fillfactor.  The
biasing and the readout of the data produced by the sensors is done by
by flat band cables located on top of the chips. The cables ship the
data to some front end boards located on top of the heat sink. Those
front end boards contain directly cooled low voltage regulators and a
multiplexer logic. The latter is to concentrate the data on a minimum
amount of optical or copper lines, which are passed through a vacuum
window to the outside world.

\subsection{Estimated material budget of the cooling support}

\begin{figure}[b]
  \begin{center}   
    \begin{minipage}[c]{7.5cm}~\\\begin{center}
      \includegraphics[width=4.5cm]{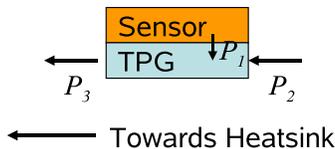}
      \end{center}
    \end{minipage}
    \begin{minipage}[c]{8cm}
      \caption[] {\it "Finite element" of the cooling support of the
        ladder. The heat streams used in the calculation are shown.
        \label{HeatElement}}
    \end{minipage}
  \end{center}   
 \end{figure}

The material budget of this detector ladder varies as function of the
cooling power required as the cooling needs determine the necessary
thickness of the TPG-layers.  As starting point for the calculation of
their thickness, we derive the temperature difference at a small
element of the latter. The length of this volume along the latter is
given with $L_V$. Its surface toward the neighboring element derived
like $S_V=W \cdot \tau_V$ from the width $W$ of the ladder and its
thickness $\tau_V$. We assume now, that this element is crossed by
three different heat flows (see figure \ref{HeatElement}): $P_1$ is
the heat injected by the chips mounted on the volume element. $P_2$ is
the heat flow, the element receives from its neighbors located
upstream the latter. $P_3$ is the heat flow, the element sends to its
neighbor directed toward the heat sink. For reasons of energy
conservation, those heat flows fulfill the equation:
\begin{equation}
P_3=P_1 + P_2
\end{equation}
A temperature gradient is required in order to drive the heat flow
through our element. For very small volume elements in the middle of
the ladder, it is justified to state that $P_2 \gg P_1$. For the
calculation of the heat flow, we can therefore approximate that $P_3
\approx P_2$. The temperature difference $\Delta T_{V_i}$ on the
volume number $i$ is then derived according to the equation of heat
conduction:
\begin{eqnarray}
P_{3_i} = \frac{\lambda}{L_V} \cdot S_V \cdot \Delta T_{V_i} \\
\Rightarrow \Delta T_{V_i} = \frac{P_{3_i} \cdot L_V}{\lambda \cdot S_V}
\label{EqnTemperatureDropElement1}
\end{eqnarray}
In this equation, $\lambda$ stands for the heat conductivity of the
material.

In order to estimate the temperature drop on the full cooling support,
one assumes that the volume element considered is the $i^{th}$ element
in a chain of equal volume elements. The element number zero is
situated at the border of the latter, which is located opposite to the
heat sink. As this element has no neighbors, $P_{2_0}=0$. For the heat
flow through the $i^{th}$ element one derives then:
\begin{equation}
P_{3_i} = i \cdot P_1
\end{equation}
Knowing the heat density $\rho_{Power}$ produced by the MAPS detectors
(according to equation \ref{EquationEnergieDichte}), one may derive
$P_1$ from the dimension of the interface between the volume element
of the sensor, which is given with $L_V \cdot W$:
\begin{equation}
P_1= \rho_{Power} \cdot L_V \cdot W = \frac{{P_{_{Block}(f)}} \cdot L_V \cdot W}{f \cdot t_{Int} \cdot P^2}
\end{equation}
With this we conclude for $P_{3_i}$:
\begin{equation}
P_{3_i} = i \cdot \rho_{Power} \cdot L_V \cdot W
\end{equation}
Combining this information with equation \ref{EqnTemperatureDropElement1}, one obtains:
\begin{equation}
\Delta T_{V_i} = i \cdot \frac{ \rho_{Power} \cdot {L_V}^2 \cdot W }{\lambda \cdot S_V}
\end{equation}
The temperature drop on the full ladder with a length of $L$ is
derived by summing up the temperature drops on the individual volume
elements. This is done like
\begin{equation}
\Delta T= \sum \limits^{L/L_V}_{i=0} \Delta T_{V_i} = \sum \limits^{L/L_V}_{i=0} i \cdot \frac{ \rho_{Power} \cdot {L_V}^2 \cdot W }{\lambda \cdot S_V}
\label{EqnSummedTemperatures}
\end{equation}
Knowing that 
\begin{equation}
\sum \limits^{N}_{i=0} ~~ i ~~ = \frac{N (1+N)}{2} \approx \frac{N^2}{2}
\end{equation}
one simplifies equation \ref{EqnSummedTemperatures} to:
\begin{equation}
\Delta T= \frac{1}{2} \cdot \frac{ \rho_{Power} \cdot {L}^2 \cdot W }{\lambda \cdot S_V}
\end{equation}
As $S_V$ was defined like $S_V=W \cdot \tau_V$, this translates into:
\begin{eqnarray}
\Delta T &=& \frac{1}{2} \cdot \frac{ \rho_{Power} \cdot {L}^2}{\lambda \cdot \tau_V} \\
\Rightarrow \tau_V &=&\frac{1}{2} \cdot \frac{ \rho_{Power} \cdot {L}^2}{\lambda \cdot \Delta T}
\end{eqnarray}
Putting the energy density $\rho_{Power}$ according to equation \ref{MapsBasisGleichung} and \ref{MapsBasisGleichung2}, one obtains:
\begin{eqnarray}
\tau_V &=& \frac{0.018}{\rm m^2 }\cdot \frac{ {L}^2}{\lambda \cdot \Delta T} \cdot \frac{ P_{col}(f)  }
{ t_{int} \cdot f } \cdot  \left(\frac{T_{non-io}}{~\rm 
n_{eq}/cm^2} \right)^{0.88} \label{DickeGleichung}\\
\Leftrightarrow
\tau_V &=&\frac{1}{2} \cdot \frac{  {L}^2}{\lambda \cdot \Delta T}\cdot \frac{P_{col}(f)}{t_{int} \cdot f \cdot (5 \sigma)^2}
\end{eqnarray}
This equation defines the necessary thickness of the cooling layer as
function of the detector requirements\footnote{Note that restrictions
  for the validity of equation \ref{radhardEquation} applies also to
  equation \ref{DickeGleichung}.}. The radiation length of this layer
can be derived knowing that TPG is a flavor of graphite.

For the calculation of the thickness of the cooling layer, we assume
as a plausible scenario that the pixel detector will operate at a
position of $z=5~\rm cm$ ($\Rightarrow L=\rm 2~cm$) with a pixel pitch
of $10~\mu m$ and a time resolution of \mbox{$20~\mu s$}. The power
dissipation of an end of column block of a MAPS
detector\footnote{Estimate for sensors manufactured in the AMS 0.35
  $\mu$m Opto process. From \cite{ChristineHu}.} is set to
$P_{col}=600~ \rm mW$. A temperature gradient of $\Delta T=15 \rm ~K$
is accepted in order to drive this heat toward the heat sink. Using
this input, we compute a $\tau_V=250~\rm \mu m$.

\subsection{Material budget of a detector station}
\begin{figure}[t]
  \begin{center}   
    \begin{minipage}[c]{9.0cm}~\\
      \includegraphics[width=8cm]{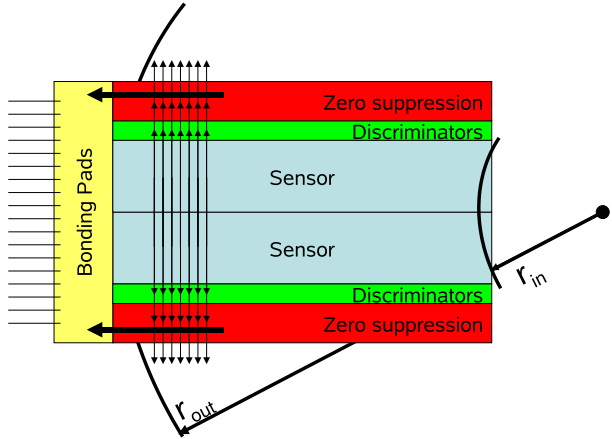}
    \end{minipage}
    \begin{minipage}[c]{7cm}
      \caption[] {\it The sensor structure proposed for a use in the MVD. The
        surface of the station is covered by a large number of individually
        short readout columns. The bonding pads, and thus the cables, are
        located outside the detector acceptance.
        \label{MVD-Sensor}}
    \end{minipage}
  \end{center}
\end{figure}

For the design of the crucial first detector station, one may profit
from the small dimensions of this station. The acceptance of CBM
requires the station at $z=5~\rm cm$ to approximate a circle an outer
radius of $r_{out}=2.5~\rm cm$ and an inner hole\footnote{This inner
  hole is slightly enlarged with respect to the global inner
  acceptance angle of CBM in order to generate the necessary room for
  the beam and to reduce the radiation doses on the station.} with a
radius of $r_{in}=0.5~\rm cm$. The length of a detector ladder is
therefore $L=2\rm~cm$ and thus slightly below the typical size of a
reticle used for designing a MAPS chip. This should allow to cover
this station with sensors having the geometrical form shown in figure
\ref{MVD-Sensor}. This geometry combines the short columns required
for a good time resolution with a width, which is sufficient for
covering the surface of the MVD station. The bonding pads of the
sensors (and the corresponding cables) are located outside the
detector acceptance. The cables do therefore not contribute to the
material budget of the system.

The remaining material budget is contributed by the \mbox{$50~\mu \rm
  m$} thick silicon of the sensors \cite{Mimosa5Thinning}, the TPG of
the cooling layers, four layers of glue and the RVC carbon foam used
to stiffen the structure. An estimate of the corresponding material
budget is shown in table \ref{TableMaterialBudget}, which suggests
that a material budget of \mbox{0.3\% $X_0$} is not out of scope for
the crucial first station of the MVD.

\begin{table} [tb]
\centering
\small
\begin{tabular}	{||c||c||r|r|r||}
		\hline
		\hline
 {\bf Material} &  {\bf Functionality} &{\bf $X_0 $ [cm] } & {\bf Thickness}& {\bf x [\%] } \\
		\hline
		\hline
{Silicon } & {Sensors }&{9.4} & {$2\times 50 ~\rm \mu m$} & {$ 0.11 $}  \\
		\hline
{TPG (Cooling)} &{Cooling} & {19.0}& {$\sim 250~\rm \mu m$} & {$\sim 0.13$}  \\
		\hline
{RVC}&{Mech. support} & {723.7} & {$3 ~\rm mm$} & {$0.04$} \\
		\hline
{Glue} & {Integration} & {$\sim$ 35.0} & {$4 \times 30~\rm \mu m$}& {$\sim 0.04$} \\
\hline \hline
	{Sum} & {Station} & {--} & {$3.5~\rm mm$}& {$\sim 0.31$} \\	
		\hline
		\hline

		\end{tabular}
\caption[Performances of different pixel detectors]{ \it Estimated
  material budget of a vertex detector station located at a position
  of \mbox{$z=5~\rm cm$}. The flat band cables used for readout do not
  contribute as they can be installed outside of the detector
  acceptance. Note that the radiation length of the glue and the
  thickness of the TPG depend on future technology choices.  }
\label{TableMaterialBudget}				
\end{table}

\section{The physics potential of the detector concept}
\label{SectionSimulation}
\subsection{Running scenario}

Newly arising pixel detector technologies like SOI-detectors or pixel
detectors based on 3D-VLSI have the potential to improve the limits of
the pixel detector technology substantially within the next
decade. However, experience shows that it needs a substantial amount
of time to evolve a promising technology into a running detector. We
suppose therefore that the promising next generation pixel sensors
will become available only after the the start of CBM and foresee
therefore two MVD generations.

The first MVD generation will rely on MAPS. Due to the existing
experience with this technology, it will be available from the start
of CBM. Given the limits of MAPS in terms of time resolution and
radiation hardness, this detector will presumably not allow to cover
the full physics program of CBM. However, it will open the door to the
so far unknown world of open charm produced in p-p and A-A collisions
at the SIS-300 top energies and thus allow for valuable physics
programs. An upgraded MVD based on next generation sensors will
presumably allow to complete the mission of CBM by measuring open
charm also at lower beam energies.

We suppose that the first generation MVD will operate with a time
resolution slightly slower than \mbox{10 $\mu$s} in order to relax the
requirements on cooling and material budget. A modest pile up of
nuclear collisions should be tolerable and allows for a collision rate
in the order of few \mbox{$10^5$} collisions per second. At this
collision rate, the radiation tolerance of MAPS is sufficient for a
reasonable operation time in the order of months. The option to
produce MAPS in cheap industrial mass production will allow for a
regular replacement of the small MVD stations, which further increases
the physics potential of the system. To limit the radiation damage in
the detector, the MVD is removed whenever CBM operates on physics
cases like di-muon spectroscopy, which do not need vertex information
but require very high beam intensities.

The trigger system for open charm will rely on a real time tracking
and on applying selection criteria on the impact parameter of
individual tracks and on the secondary vertex on the fly. Higher level
analysis may moreover use the hadron identification information from
the time of flight system of CBM. The latter allows for a $\pm
2\sigma$ separation pions and kaons with a momentum of up to $p=3.5\rm
~GeV/c$. Moreover, it may identify protons with good efficiency for $p
\lesssim 6\rm ~GeV/c$\cite{TechnicalStatusReport}, which covers most
of the momentum range of interest.

\subsection{Simulation of the physics performance}

Various simulations were performed in order to estimate the physics
performances of the MVD detector in the above discussed running
scenario. The physic simulations typically relied on a open charm
production with a thermal model and a generation of the underlying
nuclear collisions with UrQMD. Event mixing of the nuclear collisions
was required to reach the extremely high background statistics needed
for testing the efficiency of our selection criteria.

The track finding and track fitting in the simulations was done using
a cellular automaton track finder and a Kalmann filter. Both software
packages, which are currently being optimized for multi-core
processing architectures, are part of the CBMROOT simulation and
analysis package. The main selection criteria applied were the impact
parameter cut of the individual tracks and a cut on the secondary
vertex position of the decay candidate. Moreover, we checked if the
momentum vector of the reconstructed particle points to the primary
vertex.  Pile-up and $\delta$-electrons were so far neglected as the
software tools required for simulating both effects became available
only recently. The results presented hold therefore under the
assumption that an efficient track finding is possible despite the
high detector occupancies.

\begin{figure}[t]
  \begin{center}   
    \begin{minipage}[c]{10cm}~\\~\\~\\
      \includegraphics[viewport=1cm 0cm 11cm 6cm ,clip,width=8cm]{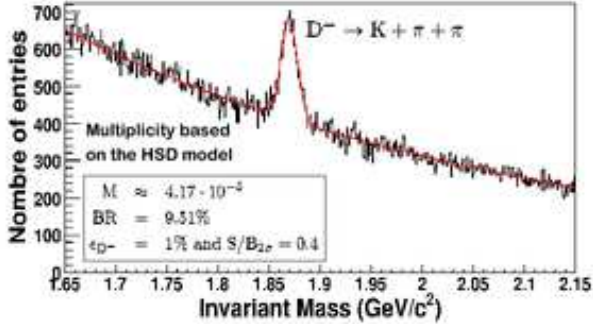}
    \end{minipage}
    \begin{minipage}[c]{5.5cm}
      \caption[] {\it The peak of a reconstructed decay of \mbox{$D^+
          \rightarrow K^- + \pi^+ + \pi^+$} from a Au+Au collision at 25
        AGeV. The amount of data shown corresponds to the lifetime of one
        set of MAPS sensors.
        \label{OpenCharmPeak}}
    \end{minipage}
  \end{center}  
 \end{figure}

The primary goal of our simulations is to optimize the design of our
detector and to understand the consequences of different technology
choices. The precise results of the simulations vary therefore
depending on the precise assumptions made on the material budget and
the lifetime of the MVD. A conservatively chosen example of a
simulation result for the reaction \mbox{$\rm D^+ \rightarrow K^- +
  \pi^+ + \pi^+$} is shown in figure \ref{OpenCharmPeak}. This
simulation assumes a relatively low\footnote{In the sense of our
  running scenario.} beam energy of \mbox{$25~\rm AGeV$} and a
material budget of \mbox{$0.3\% ~X_0$} for a station located at
\mbox{$z=10~\rm cm$}, which combines a low production multiplicity for
open charm with a modest secondary vertex resolution of \mbox{$\sim
  80~\rm \mu m$}. Nevertheless, within the lifetime of an individual
set of sensors, 5000 $\rm D^+$ mesons are reconstructed with a
$S/B=0.4$. Those numbers are expected to increase by roughly one order
of magnitude for SIS-300 top energies. Similar results were achieved
for $\rm D^0 \rightarrow K + \pi$ and for the four-body decay
\mbox{$\rm D^0 \rightarrow K^- + \pi^+ + \pi^+ + \pi^-$}. Despite its
extremely small lifetime, the $\Lambda_{\rm C}$ baryon will presumably
be visible at high beam energies. However, the number of reconstructed
particles will remain modest and allow only for measurements particle
yields. Nevertheless, the results seem sufficient to cover a
substantial part of the physics program of CBM.

\section{Summary and Conclusion}

In this work, we introduced conceptual considerations for the micro
vertex detector (MVD) of the Compressed Baryonic Matter Experiment
(CBM). The CBM experiment is a fixed target experiment. The energy of
its heavy ion beam of \mbox{8 - 45 AGeV} is optimized to study the
phase diagram of hadronic matter in the region of highest net baryon
densities. The experiment aims to find for the expected first order
phase transition of hadronic matter, signatures of chiral symmetry
restoration and the critical endpoint of the phase diagram.

The aim of the MVD of CBM is to measure the production multiplicity
and flow of open charm particles, which are so far unknown in this
energy region. The particles will be reconstructed via their hadronic
decay channels by identifying their secondary decay vertex. The latter
sets unprecedented requirements on the performance of the detector. We
motivated that an ideal MVD would have to provide a combination of
very good spatial resolution (\mbox{few $\mu$m}), light material
budget (\mbox{few $0.1\%~X_0$}) and radiation hardness of (\mbox{few
  $10^{15} \rm n_{eq}/cm^2$} + \mbox{$ 340 \rm Mrad$} per year. A time
resolution of \mbox{$ \lesssim 100~\rm ns$} is required to separate
individual nuclear collisions at the CBM top collision rate of
\mbox{$\sim 10^7$ collisions/s}.

The full set of requirements of the CBM experiment is not matched by
any existing sensor technology. In order to approach them to the
limits of nowadays technology, we chose the CMOS Monolithic Active
Pixel sensors developed at IPHC Strasbourg as our non-exclusive guide
line sensor technology. This was done as MAPS provide the necessary
spatial resolution and light material budget together with an advanced
radiation hardness of \mbox{$\sim 10^{13} \rm n_{eq}/cm^2$}. Concepts
to reach a time resolution of \mbox{$\sim 10~\mu s$} were discussed.

Based on those numbers, we introduced a design concept for an MVD
based on MAPS and discussed its features and limits. We showed
simulation results which suggest, that the concept will allow for
doing open charm physics at the CBM top energies with a reduced
collision rate of few \mbox{$10^5$ collisions} per second. This will
be sufficient for for measuring the production multiplicity for the
$\rm D^0, D^\pm$-mesons and presumably for the $\Lambda_C$. Moreover,
flow measurements might be possible for the open charm mesons. Those
abilities make CBM an experiment with unique physics potential.

The CBM collaboration is observing closely the progress in the
development of next generation pixel detectors based on novel
technologies like the SOI-sensors or sensors relying on the 3D-VLSI
integration. We envisage an upgrade of the MVD, once those promising
sensors are available. The upgrade of this detector should expand its
abilities sufficiently to complete the physics program of CBM also in
the region of lower beam energies.

\section*{Acknowledgments}

This work was supported by the 
BMBF (06FY1731), the GSI and the
Hessian LOEWE initiative through the
Helmholtz International Center for FAIR (HIC for FAIR).

\end{document}